# Quantitative Fabrication-Error Reduction in Optomechanical Crystal using Proximity Error Correction

**Pratip Ghosh[1], Anohita Mallick[2] and Akshay K. Naik[1*]**
[1]*Centre for Nano Science and Engineering, Indian Institute of Science, Bangalore-560012, India*
[2] *Department of Astronomy and Astrophysics, Tata Institute of Fundamental Research, Mumbai-400005, India*
**Corresponding author: anaik@iisc.ac.in*

**Abstract:** We demonstrate the use of proximity effect correction (PEC) in electron beam lithography (EBL) to improve the fabrication fidelity of dense photonic crystal structures. Monte Carlo simulations were employed to model electron scattering and determine the proximity function of the resist-substrate system. Based on this, a computational dose-modification scheme was implemented to compensate for nonuniform energy deposition during exposure. In addition, SEM-based image analysis was performed to quantitatively assess structural differences among uniformly exposed, manually dose-modified, and PEC-fabricated devices by comparing extracted geometries with the reference GDS design. The analysis revealed reduced dimensional deviation, improved spatial uniformity, and lower edge roughness in the PEC-corrected structures. These improvements resulted in a significantly enhanced optical quality factor in the fabricated photonic crystal cavities.

## I. INTRODUCTION

Photonic integrated circuits (PICs) on Silicon-on-Insulator (SOI) platforms have emerged as a critical technology for a wide range of applications, including data processing, communication, quantum photonics, nonlinear optics, sensing, and cavity optomechanics [1], [2], [3], [4]. These platforms enable strong optical confinement and manipulation of light within compact on-chip structures [5], [6]. Among the various microcavity architectures, photonic crystal cavities have attracted significant interest due to their ultra-small mode volumes and strong light confinement arising from the formation of a photonic bandgap [5], [7], [8], [9], [10], [11]. When these optical structures are combined with engineered mechanical resonances in suspended nanobeam geometries, optomechanical crystal cavities, that simultaneously confine photons and phonons, can be realized[4], [8], [12], [13], [14].

The performance of such devices relies heavily on nanometre-scale control of critical dimensions and feature placement. Electron Beam Lithography (EBL) is one of the key fabrication techniques capable of meeting these requirements, as it uses a focused electron beam to transfer patterns onto an electron-sensitive resist coated on the substrate with high spatial precision[15], [16], [17], [18]. However, electron scattering within the resist and substrate can cause unintended exposure of nearby regions, leading to the well-known proximity effect[19], [20], [21]. This effect can result in pattern broadening, dimensional inaccuracies, and deviations from the actual geometries.

In densely patterned structures such as optomechanical crystals, proximity effects become particularly severe due to the close spacing and subwavelength dimensions of the features[22], [23], [24], [25]. These fabrication imperfections can introduce surface roughness, hole-size nonuniformity, and structural disorder. All of these degrade the device's performance by increasing scattering losses and reducing the optical quality factor.

To mitigate these issues, computational proximity effect correction (PEC) methods can be employed to estimate the deposited energy distribution in the resist and substrate and, accordingly, optimize the local exposure dose. By selecting an appropriate proximity function and applying dose correction at specific locations in high-density designs, pattern fidelity can be significantly improved [19], [26], [27], [28], [29]. Although PEC can reduce proximity-induced fabrication errors, its effectiveness must be assessed quantitatively by comparing the fabricated geometry with the intended design. While SEM imaging provides direct visualization of the fabricated structures, extracting dimensional and positional deviations from dense photonic-crystal geometries is non-trivial. A quantitative SEM-to-design comparison is therefore required to determine how PEC modifies fabrication fidelity and, ultimately, the device performance.
In this work, we combine Monte Carlo simulation-based PEC with a Python-based SEM analysis workflow to quantitatively evaluate fabrication fidelity in optomechanical crystal structures[30]. The analysis enables comparison of the fabricated geometry with the intended GDS design and extraction of geometric deviations under different exposure conditions. We further correlate these fabrication characteristics with the devices' optical performance, demonstrating an enhancement in the optical quality factor following PEC.

## II. PROXIMITY EFFECT AND PEC METHOD

To understand the origin of the proximity effect and the basis of the Monte Carlo-based correction used in this work, we first consider the electron-scattering processes that govern energy deposition during EBL. When a high-energy electron beam (typically in the keV range) interacts with the resist, the incident electrons undergo multiple scattering events, continuously losing energy and changing their trajectories as they penetrate the resist and substrate[17], [19], [20]. This process is referred to as electron scattering. During these interactions, incident electrons can break the molecular chains of the resist via energy transfer, thereby modifying its solubility during development[20], [23]. Electron scattering in EBL is generally classified into two main types: forward scattering and backscattering.
(a) Forward scattering occurs when the incident electrons pass through the resist with only small-angle deflections. This is mainly associated with interactions within the resist and contributes to a slight broadening of the beam profile.
(b) Backscattering occurs when electrons penetrate deeper and undergo large-angle elastic scattering. These are primarily due to interactions with the atomic nuclei in the substrate. These electrons retain a significant fraction of their energy and can travel back toward the resist, exposing regions away from the intended pattern.
In addition to primary and backscattered electrons, secondary electrons are generated by inelastic energy-loss processes. Although they have relatively low energy (on the order of a few keV), they play an important role in resist exposure at the local scale. The combined effect of forward-scattered, backscattered, and secondary electrons results in an unintended spatial distribution of deposited energy, giving rise to the proximity effect in electron beam lithography [21], [23].
Figure 1(a) illustrates the electron scattering model, showing the interaction of incident electrons with both the atomic electrons and the nuclei of the material. Figure 1(b) presents a schematic representation of the incident electron beam along with forward-scattered and backscattered electron trajectories. The beam diameter at the bottom of the resist can be described by[18], $d =$ $0.9\left(\frac{t}{V}\right)^{1.5}$ nm; where $t$ represents the resist thickness in nm and $V$ denotes the acceleration voltage in kV. This expression suggests that reducing the resist thickness and increasing the beam acceleration voltage led to a smaller beam diameter at the resist-substrate interface, thereby minimizing forward scattering effects. In contrast, the range of backscattered electrons is primarily influenced by the atomic numbers of the resist and substrate materials and the incident beam energy [31].

Materials with higher atomic numbers generally exhibit a larger backscattering coefficient, resulting in a greater lateral spread of backscattered electrons and consequently a wider unintended exposure area. Likewise, increasing the acceleration voltage enables deeper electron penetration into the substrate, which can further increase the lateral extent of backscattered electrons.

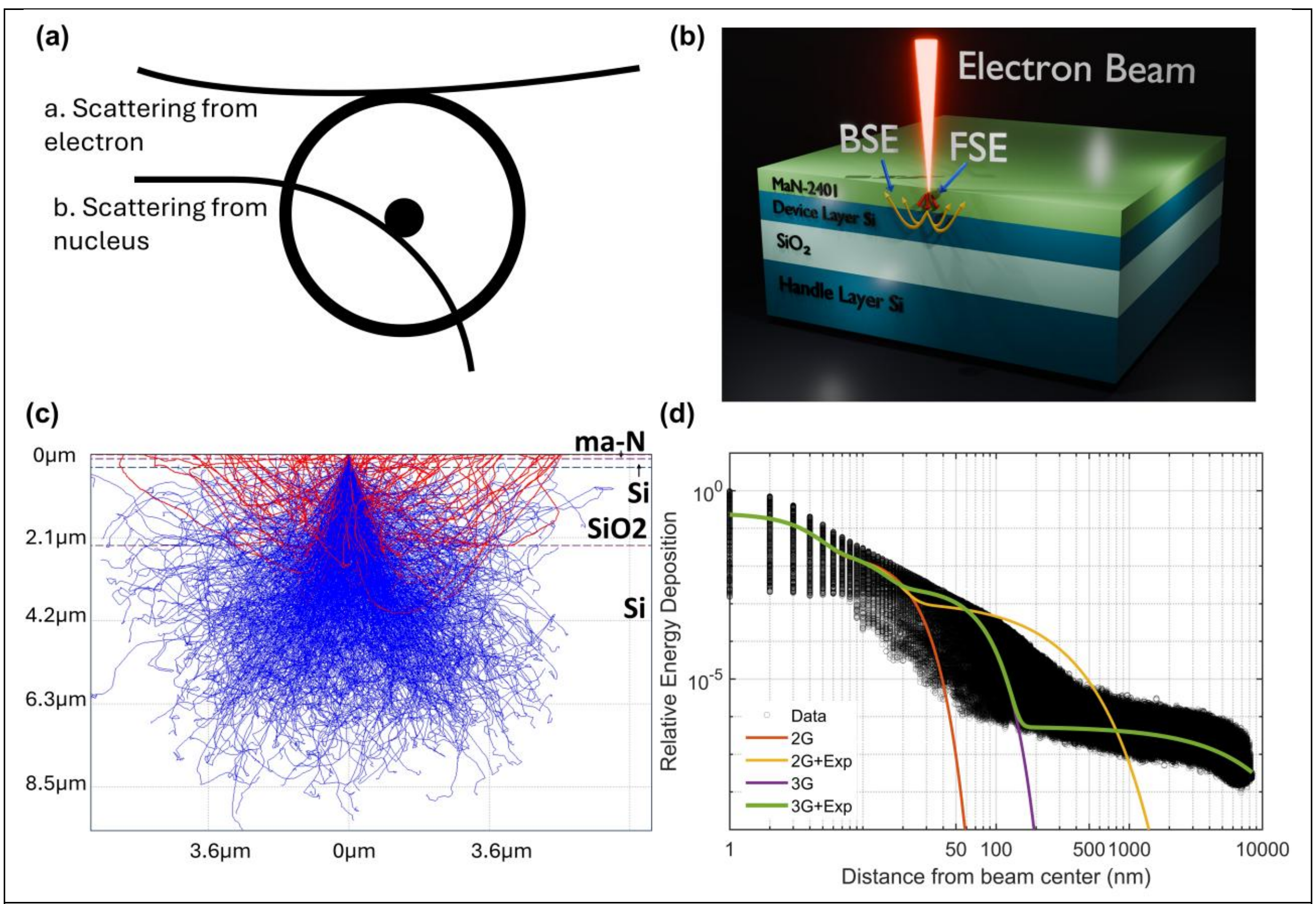

***Figure 1: (a)**The electron scattering model for incident electrons with atom. (**b)** schematic of electron beam with Forward and backward scattering electrons. **(c)** Scattering trajectories of electrons from resist and substrate. **(d)** Fitted data with different profiles of the system*

Several methods have been developed for proximity effect correction, including shape modification, dose modification, and background-corrected exposure (commonly known as GHOST) [27], [28], [29]. Each of these techniques offers certain advantages and limitations depending on the pattern geometry and fabrication requirements. In this work, we employ the dose modification approach, in which the exposure dose is optimized locally for each pixel. Because the required dose varies with the surrounding pattern density, these pixel-wise dose values are determined through computational calculations.

After EBL exposure, the final resist profile is formed during the development process and is governed by the local dose deposited during exposure. For negative resists, regions exceeding a dose threshold become less soluble and remain intact, while underdosed areas dissolve completely. Accurate profile prediction requires identifying the dose for complete resist retention ($D_{100}$) and the minimum dose for full dissolution ($D_0$)[23]. The deposited dose influences the degree of cross-linking in the resist, which in turn affects its solubility, critical dimensions, and the final development profile.

To model these effects, the spatial map of energy deposited by a point-source electron beam in the resist-substrate stack must be computed. This map, called the Energy Intensity Distribution (EID), underpins local dose evaluation across patterns and enables correction for proximity effects. Here, EID profiles were generated via Monte Carlo simulations of electron scattering during EBL[30]. Figure 1(c) presents the Monte Carlo simulated trajectories of forward-scattered and backscattered electrons in the MaN-2401/SOI resist-substrate stack for an acceleration voltage of 30 keV. The simulation was carried out using CASINO with 10,000 incident electrons.

Optomechanical crystal designs are typically densely packed. The exposure received by a given feature is influenced by the direct incident beam and by the energy deposited from neighbouring regions due to the proximity effect. Thus, a uniform exposure dose can still lead to nonuniform cross-linking and distortions of the structure's critical dimensions. To address this, the design can be discretized into $N$ pixels[32]. As shown in Fig. 1(d), a suitable proximity function can then be selected to describe the Energy Intensity Distribution (EID) within the resist. In this work, we have modelled the proximity function as the sum of three Gaussian components and one exponential term. The total deposited energy at the $i$-th pixel, $E_i$, is then given by

$$E_i = \sum_{j=1}^{N} P_{ij} D_j \tag{1}$$

where $D_j$ is the applied dose at the $j$-th pixel and $P_{ij}$ represents the proximity interaction between pixels $i$ and $j$. This method is known as a self-consistent scheme[27], [28], [33]. Its main limitation, however, is the high computational time required to solve the dose distribution for complex layouts, as the number of pixels increases significantly with design density and size.

## III. DEVICE FABRICATION AND OPTICAL CHARACTERIZATION

An SOI substrate coated with 120-nm MaN negative e-beam resist was used to fabricate the devices. The patterns were exposed using an electron beam energy of 30 keV with a 10 µm aperture. The design consists of a dense nanophotonic structure with critical feature sizes down to 70 nm. Figure 2(a) shows an SEM image of the fabricated optomechanical crystal device, comprising a photonic crystal cavity, a phononic shield, and a bus waveguide for optical coupling to the cavity. A key fabrication objective was to accurately define the elliptical cavity features while maintaining low edge roughness. When a uniform exposure dose was applied, incomplete development was observed in the corner regions (red rectangle in Figure 2), and the elliptical and circular features in the cavity region (see blue rectangle in Figure 2) were not well resolved. In contrast, the device shown in Figure 2(b) was fabricated using proximity-effect correction (PEC) via dose modification, resulting in smoother edges and a more well-defined elliptical cavity profile. To eliminate process-induced discrepancies, both devices were fabricated on the same chip and in the same exposure run. The SEM images shown correspond to the structures after development. The resist was developed using AZ 726 MIF for 26 s.

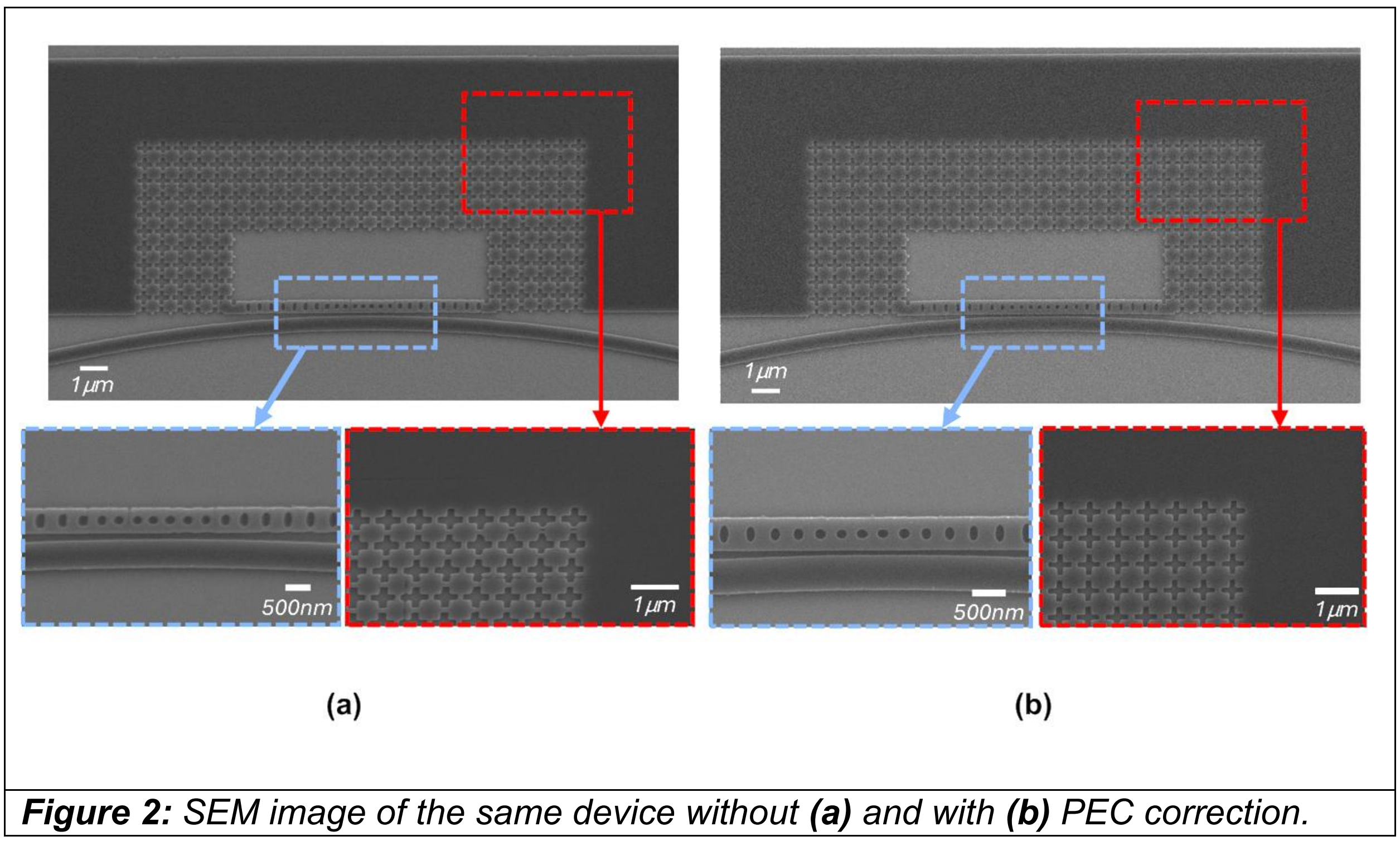

***Figure 2:** SEM image of the same device without **(a)** and with **(b)** PEC correction.*

After the simulation, we have chosen the proximity function as,

$$P_{ij} = \frac{1}{\pi(1+\eta+\nu+\nu_2)}\left[\frac{1}{\alpha^2}e^{\left(-\frac{r^2{}_{ij}}{\alpha^2}\right)} + \frac{\eta}{\beta^2}e^{\left(-\frac{r^2{}_{ij}}{\beta^2}\right)} + \frac{\eta}{2\gamma^2}e^{\left(-\frac{r_{ij}}{\gamma}\right)} + \frac{\nu_2}{{\gamma_2}^2}e^{\left(-\frac{r^2{}_{ij}}{{\gamma_2}^2}\right)}\right] \quad (2)$$

where $r_{ij}$ is the lateral distance between pixels $i$and $j$. The narrow Gaussian (width $\alpha$) captures localized energy deposition from forward scattering and short-range secondary electrons. The broader Gaussian (width $\beta$, weight $\eta$) models a wider lateral spread, while the third Gaussian (width $\gamma_2$, weight $\nu_2$) refines the fit to simulated profiles. The exponential term (decay length $\gamma$, weight $\nu$) describes the long-range backscattered tail [34], [35].

Fitted parameters from Monte Carlo-simulated EID in the MaN-2401-SOI stack at 30 keV are: $\alpha =$ 2.6 nm, $\beta = 3574.8$ nm, $\eta = 0.306$, $\gamma = 134.6$ nm, $\nu = 0.006$, $\gamma_2 = 53.98$ nm, $\nu_2 = 0.003$.

This proximity function computes total deposited energy per pixel via the self-consistent relation in equation (1). After applying the proximity-effect correction, the local exposure dose was adjusted based on the surrounding pattern density, with dose-scaling factors ranging from 0.8 to 1.6 relative to the nominal dose (Fig. 2b). This ensured that the effective energy deposited remained within the resist response window bounded by $D_0$ and $D_{100}$.

Note that the device in Fig. 3(a) was fabricated without PEC but with uniform manual dose adjustment across the cavity region to enhance pattern fidelity. This differs from the fully uncorrected structure in Fig. 2a, which received no such adjustment. Despite this manual optimization, the suspended cavity still does not preserve the intended elliptical geometry accurately and exhibits a cavity quality factor of only $\sim 3 \times 10^3$. In contrast, the PEC-corrected device shows substantially improved structural fidelity and a cavity quality factor of $\sim 3 \times 10^4$. The statistical distribution of the measured Q factors for multiple devices is shown in Fig. 3(d), demonstrating a substantial improvement in Q factor with PEC. The full fabrication process consisted of three lithography steps, two dry-etching steps, and one wet-etching step to suspend the devices.

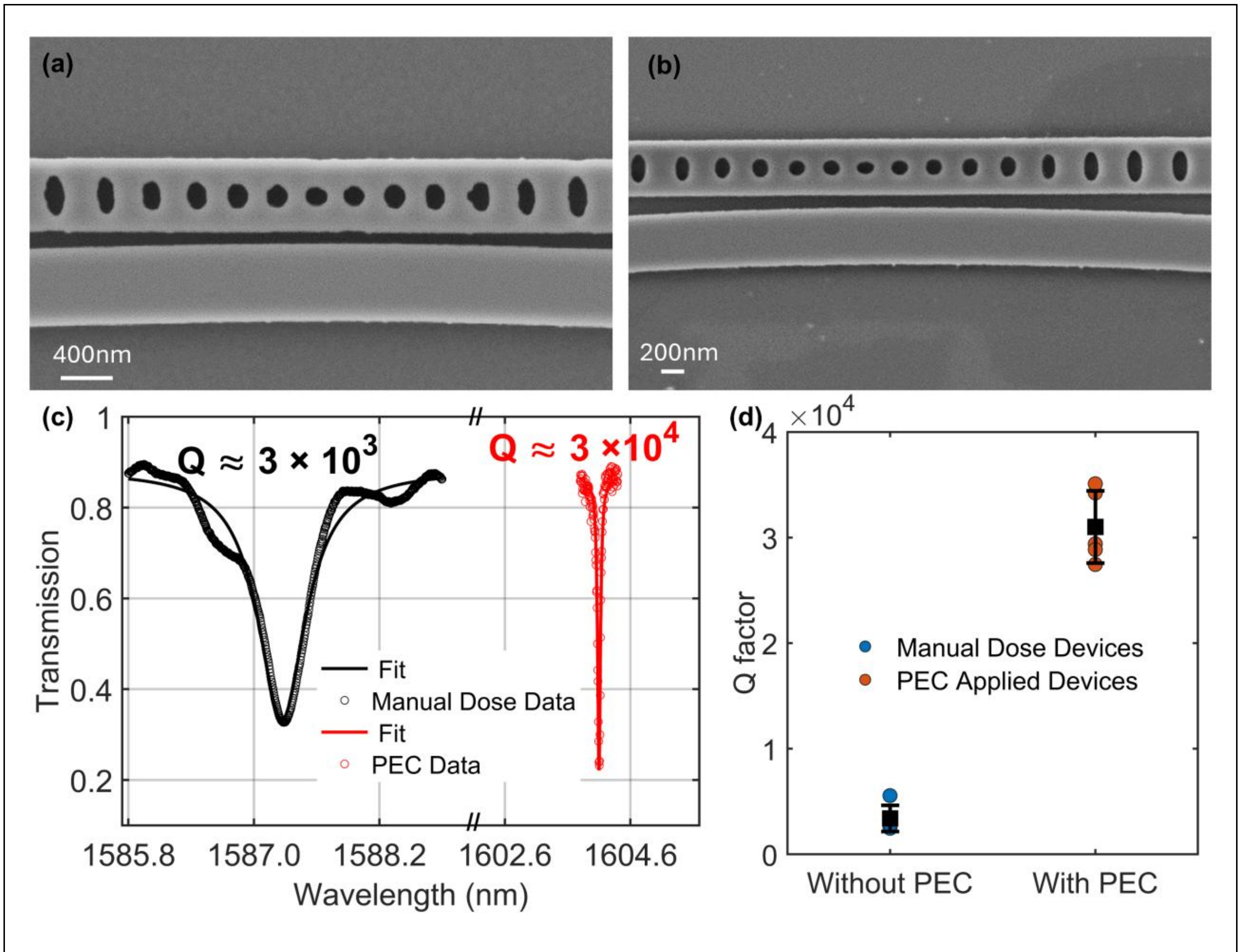

***Figure 3:*** *SEM images of the cavity region with* ***(a)*** *manual dose correction and with* ***(b)*** *PEC correction cavity.* ***(c)*** *optical transmission response for both cases.* ***(d)*** *Measured Q factors of devices fabricated with and without PEC. Points represent individual devices; filled squares indicate the mean ± standard deviation.*

While the enhancement in the optical quality factor demonstrates the benefit of PEC, a direct comparison between the fabricated geometry and the intended design is required to determine how PEC modifies the physical structure. We therefore performed a quantitative SEM-to-GDS analysis of devices fabricated under the different exposure conditions.

# IV. QUANTITATIVE SEM-TO-GDS ANALYSIS

To quantitatively evaluate fabrication improvements associated with PEC, we developed a custom Python-based SEM metrology workflow that directly compares fabricated nanostructures with the corresponding reference geometry defined in the GDS layout. The workflow was designed to operate automatically on SEM images of dense optomechanical crystal structures and consists of four main stages: (i) SEM image preprocessing, (ii) automated feature detection and contour extraction, (iii) geometric registration between SEM and GDS coordinates, and (iv) quantitative error analysis.

SEM images were first contrast-enhanced and filtered to improve the reliability of feature segmentation[36]. Device-specific morphological criteria were then used to identify the phononic-shield cross structures and the photonic-crystal cavity holes. For the cavity region, the complete contour of each detected hole was extracted and retained for subsequent shape analysis, rather than reducing the feature to fitted ellipse parameters alone. The extracted SEM feature coordinates were aligned to the GDS layout using a global affine transformation followed by nearest-neighbour feature correspondence[37], [38]. This enables direct comparison between the fabricated and intended geometries. For each feature, we have evaluated dimensional deviations in width, height, spacing, position, and elliptical radii relative to the design values.

For the photonic-crystal cavity, contour deviations were further represented as an angular radial-error function $\Delta r(\theta)$ with respect to the corresponding GDS ellipse. This contour-error signal $\Delta r(\theta)$ was decomposed into angular Fourier components. Following previous Fourier-based analyses [39], [40] of fabrication imperfections in photonic-crystal holes, the lowest-order angular modes were associated with coherent, large-scale shape deformation, whereas higher-order modes were attributed to local/contact edge roughness (see section 2C. Edge Quality and Composition of Deviations). The RMS amplitudes of the reconstructed low- and high-frequency components were then used to quantify the distortion and LER contributions, respectively.

This contour-based analysis provides a more complete assessment of fabrication errors by quantifying conventional critical-dimension measurements, such as hole width, height, and spacing, in addition to estimating the deviation of the entire fabricated hole boundary from the intended GDS shape. The modular structure of the workflow allows the same analysis framework to be applied consistently across uniformly exposed, manually dose-modified, and PEC-fabricated devices, enabling a reproducible and quantitative comparison of fabrication fidelity across different exposure conditions.

We analyzed the fabricated devices by directly comparing SEM images with the reference geometry defined in the GDS layout. The structure comprises two primary components: (i) a phononic shield formed by an array of cross-shaped ("plus") features, and (ii) a photonic crystal cavity consisting of elliptical holes embedded within a central beam. For all fabrication conditions (uniform/normal, manual, and PEC), we extracted geometric parameters from SEM images and quantitatively compared them to the design to evaluate dimensional accuracy, spatial uniformity, and edge quality. In particular, the deviations in the hole dimensions $(r_x, r_y)$ and the positional shift of the fabricated features from the intended GDS coordinates were analyzed to quantify the improvement achieved after PEC.

We calibrated the SEM images using the scale bar to obtain the pixel-to-nanometer conversion factor and aligned the extracted feature coordinates with the GDS layout using a global transformation, enabling direct comparison between fabricated and designed geometries [37]. Statistical analysis of the extracted parameters was performed to compare the fabrication fidelity for different exposure conditions. We provide detailed descriptions of the design, parsing, feature extraction, and alignment procedures in the Supplementary Information [38].

### *1. Phononic Shield: Spatial Deviation Analysis*

The phononic shield consists of a lattice of cross-shaped features arranged in rows, which provides a spatial indexing framework for quantitative analysis. For each feature, we extracted the width, arm length, and nearest-neighbour spacing (between the centres of adjacent plus-shaped features) from the SEM images and compared them with the corresponding design values obtained from the GDS layout[36]. The deviations were defined as Δlength, Δwidth, and Δspacing, and were evaluated across all fabrication conditions to quantify the dimensional variations introduced during fabrication.

In Fig. 4(a), we show a zoomed-in SEM image of two neighbouring plus-shaped structures indicating the extracted geometric parameters, length, width, and centre-to-centre spacing, used to compute the dimensional deviations. In Fig. 4(b), we show the spatial distribution of Δlength, Δwidth, and Δspacing across the lattice as continuous surfaces parameterized by row number and feature index (see Fig. S1 in the Supplementary Information). Positive and negative values indicate over- and under-sizing relative to the design, respectively. These maps reveal clear spatial trends, demonstrating that fabrication-induced deviations vary systematically across the device rather than randomly. In particular, the PEC-fabricated structures exhibit reduced deviation magnitudes and improved spatial uniformity compared to the uniform and manually dose-modified exposure conditions, indicating better dimensional control throughout the lattice.

We also observe noticeable differences in the overall uniformity and spread of the extracted parameters between fabrication conditions, highlighting the influence of the exposure strategy on feature fidelity and pattern transfer accuracy. Detailed descriptions of the feature detection, filtering, and spatial interpolation procedures are provided in the Supplementary Information.

### *2. Photonic Cavity: Geometry and Shape Analysis*

#### A. Feature Extraction and Overlay

We identified the elliptical holes forming the photonic crystal cavity in the SEM images and extracted their contours for geometric analysis. For each hole, an ellipse was fitted to determine its principal radii, orientation, and centre position. After aligning the SEM data with the GDS layout, we overlaid the extracted contours and fitted ellipses onto the reference design, as shown in Fig. 5a. This enabled a direct comparison of the fabricated cavity geometry under different exposure conditions (normal, manual dose-modified, and PEC).

The overlay provides a direct visual comparison between the fabricated and intended geometries, confirming the accuracy of the extraction and alignment procedure while clearly revealing deviations in both feature size and shape. In particular, the PEC-fabricated cavities show improved agreement with the design geometry, better preservation of the intended elliptical profile, and reduced positional deviation compared to other fabrication conditions. These observations are consistent with the improved dimensional uniformity and reduced proximity-induced distortion obtained after PEC. Detailed descriptions of the image processing, contour extraction, and ellipse fitting workflow are provided in the Supplementary Information.

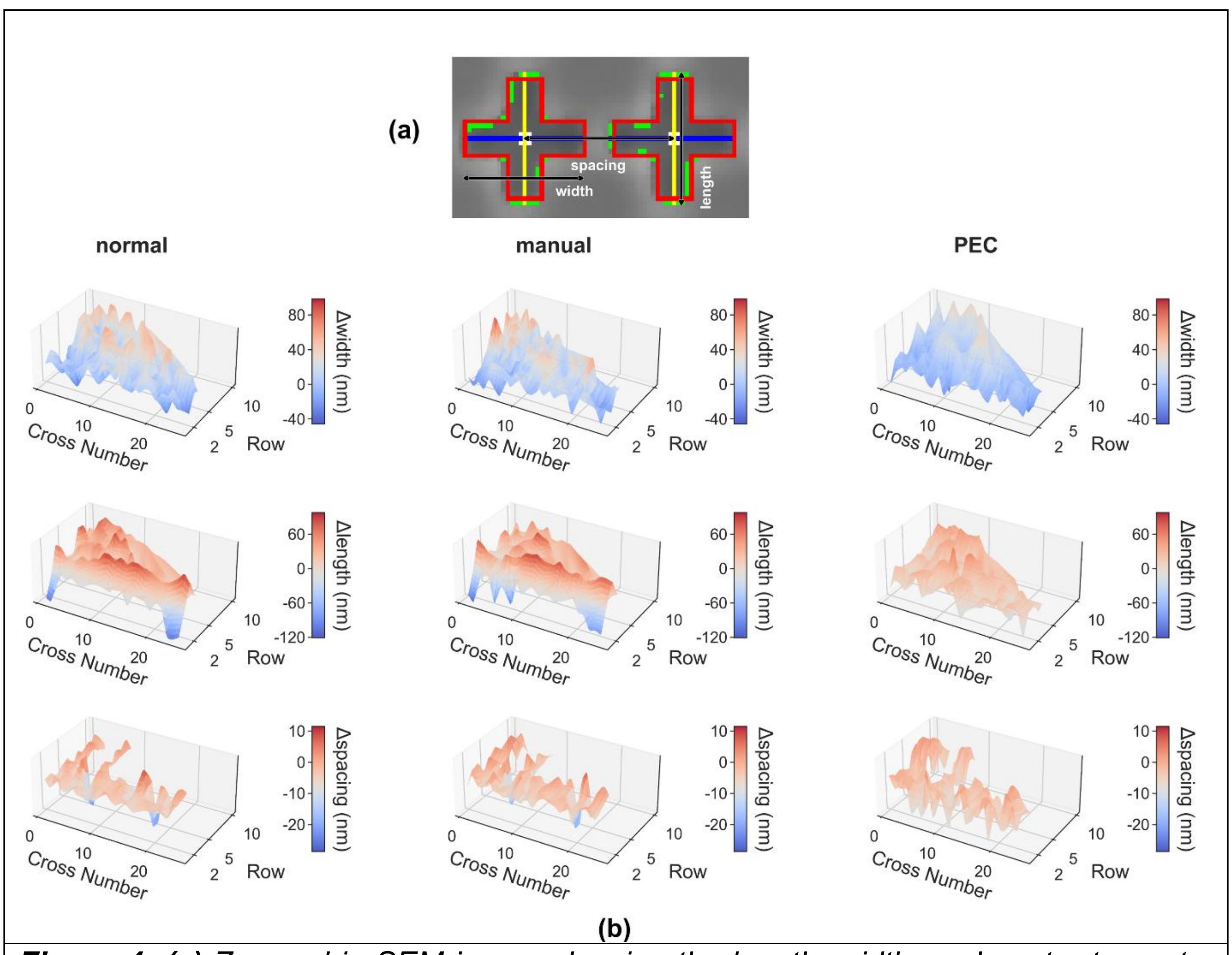

***Figure 4:*** ***(a)*** *Zoomed-in SEM image showing the length, width, and center-to-center spacing of two adjacent cross-shaped features. Green and red outlines denote the SEM-extracted contour and corresponding GDS geometry, respectively.* ***(b)*** *Spatial maps of Δlength, Δwidth, and Δspacing across the phononic shield lattice for different fabrication conditions.*

### B. Hole Size Deviations

We quantified dimensional variations in the cavity by comparing the extracted horizontal and vertical radii of each hole with the corresponding design values from the GDS layout. We plot the resulting $\Delta r_x$ and $\Delta r_y$ as a function of hole index in Fig. 5b for all fabrication conditions.

These spatial profiles reveal both systematic trends and local variations in the hole dimensions along the cavity. Clear differences between the fabrication approaches are observed in both the magnitude and uniformity of the deviations, providing insight into their effectiveness in preserving the intended geometry. In particular, the PEC-fabricated cavity exhibits reduced fluctuations and improved agreement with the design dimensions, indicating better control over feature size and placement throughout the cavity region. These results are consistent with the improved structural fidelity observed in the SEM images and the enhanced optical quality factor measured for the PEC devices.

### C. Edge Quality and Decomposition of Deviations

We further assessed our fabrication fidelity by comparing the extracted hole contours with the corresponding ideal GDS ellipses. The total deviation was quantified using the root-mean-square (RMS) error, which represents the overall discrepancy between the fabricated and designed boundaries for each cavity hole.

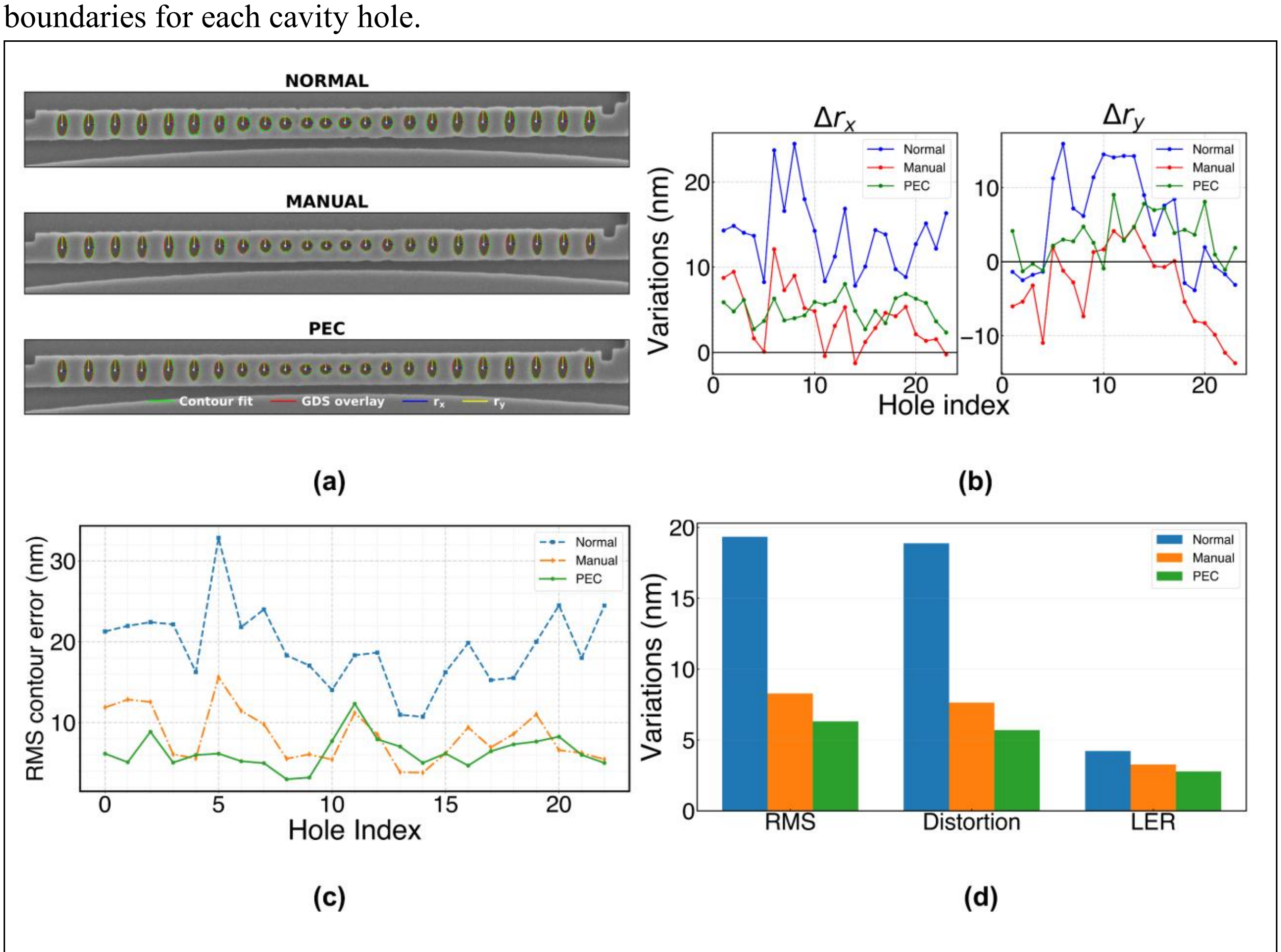

***Figure 5:*** *(a) Overlay of extracted hole contours and GDS layout. (b) $\Delta r_x$ and $\Delta r_y$ vs hole index. (c) RMS variations along the photonic cavity holes. (d) Mean RMS, distortion, and LER across holes for different fabrication conditions*

To separate systematic shape distortion from local edge roughness, $\Delta r(\theta)$was decomposed into angular Fourier modes. Modes $k \leq 4$ were taken to represent low-spatial-frequency, coherent shape distortion, whereas modes $k > 4$ were associated with higher-spatial-frequency line-edge roughness (LER). This choice follows previous studies of fabrication-induced geometrical imperfections in photonic crystal holes [39], in which the lowest few angular harmonics describe the dominant coarse-scale shape variations, while higher-order modes represent noise-like edge fluctuations. The corresponding RMS components satisfy:

$$\mathrm{RMS}^2 = \mathrm{LER}^2 + \mathrm{Distortion}^2$$

The spatial variation of the total RMS error along all holes in the cavity is plotted in Fig. 5c, providing a position-resolved view of edge quality and geometric fidelity under different fabrication conditions.

The mean values of RMS error, distortion, and LER across all holes are summarized in Fig. 5d. The PEC-fabricated devices show the lowest values for all three metrics, demonstrating a simultaneous reduction in both systematic shape distortion and local edge roughness. This reduction in both systematic and stochastic fabrication errors is consistent with the enhanced dimensional control observed in the SEM analysis and the improved optical quality factor measured for the PEC devices. Additional details of the contour-based analysis and decomposition methodology are provided in the Supplementary Information. The present implementation establishes a flexible, fully scriptable computational framework for the automated extraction and quantitative analysis of fabrication-error metrics in complex nanophotonic structures. Because the workflow integrates SEM image processing, contour-based geometric characterization, and direct comparison with the reference GDS layout within a unified Python environment, it can be readily adapted to future high-throughput SEM characterization, statistical process monitoring, and fabrication-feedback workflows.

## V. CONCLUSION

In conclusion, we demonstrate that proximity-effect correction (PEC) in electron-beam lithography (EBL) significantly improves the fabrication fidelity of dense optomechanical crystal cavities on an SOI platform. By employing Monte Carlo-derived proximity functions together with a self-consistent dose modulation scheme, we compensated for the non-uniform energy deposition arising from electron scattering in the MaN-2401/SOI resist-substrate stack.

Quantitative SEM-based image analysis, performed by direct comparison with the GDS reference geometry, revealed that the PEC-fabricated devices exhibit reduced dimensional deviations, improved spatial uniformity, lower line-edge roughness (LER), and reduced systematic shape distortion compared to uniformly exposed and manually dose-modified structures. In particular, the extracted cavity hole dimensions and contour analysis showed improved preservation of the intended elliptical geometry and more accurate feature placement throughout the cavity region.

These improvements in structural fidelity directly translated into enhanced optical performance, resulting in an optical quality factor of $Q \sim 3.5 \times 10^4$, compared to $Q \sim 3 \times 10^3$ for devices fabricated without PEC. Notably, this level of improvement was achieved despite the practical fabrication limitations associated with a 30-keV EBL system and the use of a MaN negative tone resist for dense nanoscale patterning. These results highlight the critical role of PEC in achieving high-performance nanophotonic and optomechanical devices that require precise nanoscale dimensional control. The developed SEM metrology framework further provides a basis for the automated statistical characterization of fabrication variability and for the future integration of lithographic correction with fabrication-feedback strategies.

## ACKNOWLEDGMENT

The authors acknowledge the support of the Department of Science and Technology (DST), Government of India, under the National Quantum Mission (NQM).

## Supplementary Information

### S1. Device Layout and Spatial Organization

We based the entire analysis on the GDS layout, which we used as the reference design for all geometric comparisons, from which we extracted and stored all geometric features as polygonal data [1]. The structure consists of two primary components: (i) phononic shield structures in the form of cross-shaped ("plus") features arranged in a 10-row grid, and (ii) a photonic crystal cavity composed of elliptical holes embedded within a rectangular beam.

The phononic shield is organized such that the first five rows contain 25 uniformly spaced plus structures, while in the subsequent five rows, a central rectangular exclusion region results in two separated groups of five pluses on either side. The photonic cavity is in the final row, positioned between the plus structures. This geometry defines a non-uniform spatial distribution, which we explicitly account for in the analysis. This spatial organization defines the row and feature indices used throughout the analysis and forms the basis of the spatial mapping used in subsequent visualization.

In Fig. S1, we show a representative SEM image of the full structure, where we annotate the row indices (vertical direction) and feature indices (horizontal direction). This annotated image establishes the coordinate system used to organize all extracted measurements and enables consistent comparison between SEM datasets and the reference design.

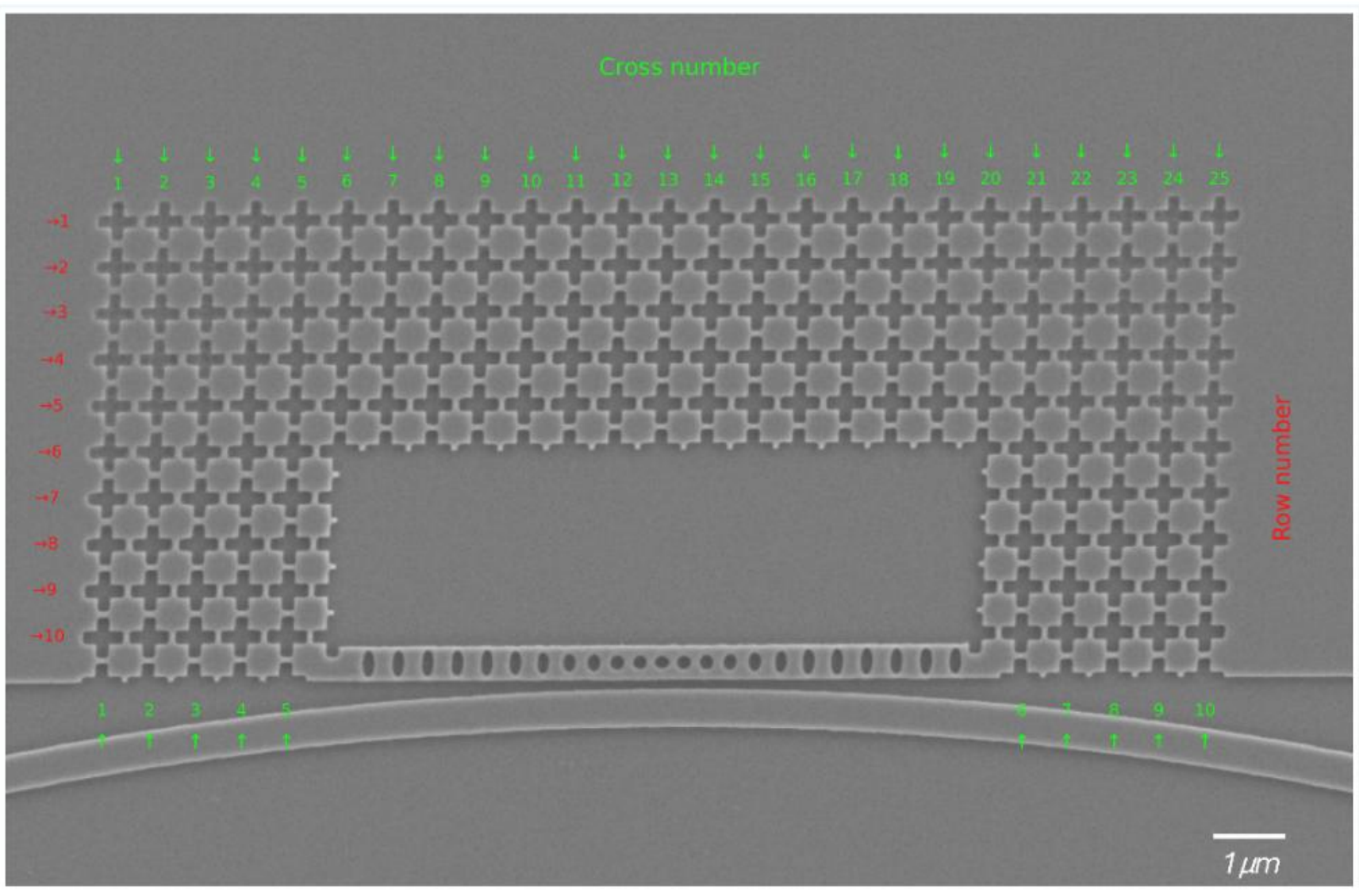

***Fig. S1*** *Sample SEM image of the device with row and feature indexing used for analysis.*

### S2. GDS Design Extraction

We extracted the reference geometry directly from the GDS layout using a Python-based workflow built on *gdstk* (GDSII Tool Kit), a C++ library with Python bindings for creating and manipulating GDSII files. We recursively traversed all hierarchical cells in the layout and collected every polygon corresponding to the device geometry.

We merged these polygons using a boolean union operation to reconstruct the complete structure. To isolate the different etched regions, we defined a rectangular bounding box slightly larger than the device and subtracted the merged geometry from it, thereby obtaining the inverse pattern.We processed each resulting polygon individually. We excluded polygons with fewer than six vertices to avoid poorly defined or degenerate shapes. For the remaining polygons, we computed geometric properties, including:

- centroid position
- bounding box (width and height)
- area
- perimeter

We calculated the circularity metric $4\pi A/P^2$to classify features. Polygons with high circularity correspond to elliptical holes, while the remaining polygons represent non-elliptical features, primarily the cross-shaped structures.

## S3. Elliptical Feature Analysis from Design

For each polygon classified as an elliptical hole, we centered the vertex coordinates about the centroid and computed the covariance matrix. We performed an eigenvalue decomposition of this covariance matrix to obtain the principal axes and the feature's orientation. We additionally estimated the semi-major and semi-minor axes from the bounding box dimensions. We converted these dimensions to nanometers and rounded them to the nearest value on the fabrication grid to reflect design discretization.

## S4. Cross-Shaped Feature Analysis from Design

For non-elliptical polygons, we analyzed the unique coordinate values along the x and y directions. We computed the smallest non-zero spacing between successive coordinates to estimate the characteristic arm thickness of each feature. We obtained the overall width and height from the bounding box. We then applied a median-based filtering criterion across all features to retain only consistent cross-shaped (“plus”) structures and remove outliers. Finally, we obtained the spatial arrangement of both elliptical and cross-shaped features by computing the nearest-neighbour distance between feature centroids. We stored all extracted parameters, including centroid position, radii, size, orientation, spacing, ellipticity, and contours, for subsequent comparison with SEM-extracted features.

## S5. SEM Image Calibration

We performed spatial calibration of each SEM image using a Python-based routine implemented in OpenCV (an open-source computer vision and image processing library, originally developed by Intel)[2]. We manually selected two endpoints of the scale bar and computed the pixel distance between them. Using the known physical length of the scale bar, we calculated a pixel-to-nanometer conversion factor, which we used for all subsequent dimensional measurements.

## S6. Detection and Analysis of Phononic Shield Features

We analyzed the cross-shaped features from SEM images using OpenCV [2]. We first smoothed the grayscale image using a Gaussian filter to suppress noise. We then converted the image into a binary representation using adaptive thresholding. We applied a mask to exclude non-relevant regions. We performed morphological operations, specifically opening followed by dilation, to remove noise and improve feature continuity.

To identify the cross-shaped features, we extracted horizontal and vertical components using directional morphological filters. We identified candidate centers of the plus structures at the intersections of these components. Around each candidate center, we analyzed a local region to isolate the connected component corresponding to a single feature. We applied geometric constraints-including limits on area, bounding box aspect ratio, and symmetry of arm lengths-to filter out spurious detections and retain only valid phononic shield elements.

In **Fig. S2**, we show the detected features for all SEM images (normal, manual, PEC). We overlay the extracted contours (green) with the corresponding GDS features (red). We also annotate the measured width and arm length for each feature (blue and yellow), enabling direct verification of the extraction process.

For each validated feature, we computed:

- centroid
- width (maximum horizontal extent)
- height (maximum vertical extent)
- arm thickness (extent perpendicular to each arm direction)

We converted all measurements to nanometers using the calibration factor. We grouped the detected features into rows based on their vertical positions and sorted them laterally. We computed inter-feature spacing along each row while excluding large gaps corresponding to the cavity region.

**S7. Alignment Between SEM and GDS Features**

To compare fabricated structures with the reference design, we aligned SEM-extracted feature coordinates with the GDS layout. We first applied a scaling transformation to account for differences in coordinate range. We then estimated an affine transformation - a linear mapping that accounts for scaling, rotation, and translation.

We used a RANSAC-based approach to robustly estimate the transformation parameters while minimizing the influence of mismatched points. We performed nearest-neighbour matching between SEM and GDS features using a k-d tree (an efficient spatial search structure) to identify corresponding points. We applied the transformation to the GDS features and overlaid them onto the SEM images, enabling direct visual and quantitative comparison.

**S8. Spatial and Statistical Analysis of Phononic Shield Deviations**

We quantified deviations in width, length (arm extent), and nearest-neighbour spacing by computing:

- $\Delta$width
- $\Delta$length
- $\Delta$spacing

as the difference between each SEM-extracted and GDS-design values.

We mapped these deviations across the phononic shield lattice using row index and feature index (see S1). We interpolated the discrete measurements onto a regular grid using linear interpolation to generate continuous spatial maps and applied a Gaussian smoothing filter to reduce noise and emphasize underlying spatial trends. We visualized these data as three-dimensional surface plots in the main text. Here, the lateral axes correspond to the feature index (along a row) and the row number, and the vertical axis represents the magnitude of deviation.

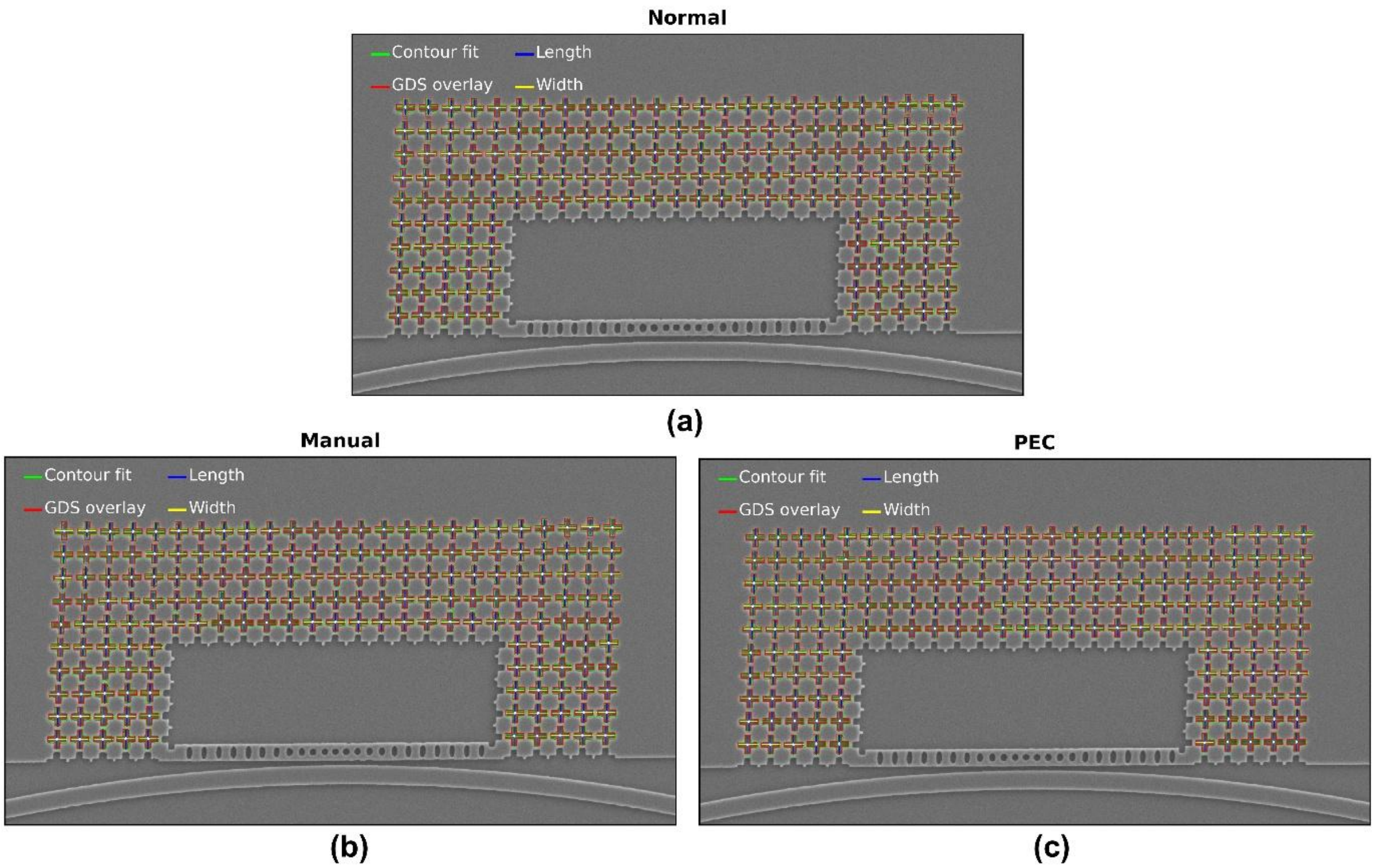

***Fig. S2*** *Detected phononic shield features with extracted contours and measured dimensions overlaid on SEM images for different fabrication conditions. Cross-shaped feature contours (green) and corresponding GDS features (red) are overlaid. Measured parameters, width and length are annotated (blue and yellow).*

In these plots, peaks and dips correspond to positive and negative deviations from the design, respectively, indicating regions where features are systematically larger or smaller than intended. The colormap provides an additional visual encoding of these variations, with contrasting colors representing the magnitude and sign of the deviation, thereby enabling intuitive identification of spatial patterns and non-uniformities across the device for different fabrication conditions (manual, uniform, and PEC). To further analyze the distributions, we generated histograms and applied kernel density estimation (KDE), which provides a smooth, continuous approximation of the underlying probability distribution.

In **Fig. S3**, we present:

- Left: KDE plots (density-normalized) for Δlength, Δwidth, and Δspacing for all fabrication conditions
- Right: histogram count plots for the same quantities

We included a zero-reference line in all plots to indicate perfect agreement with the design

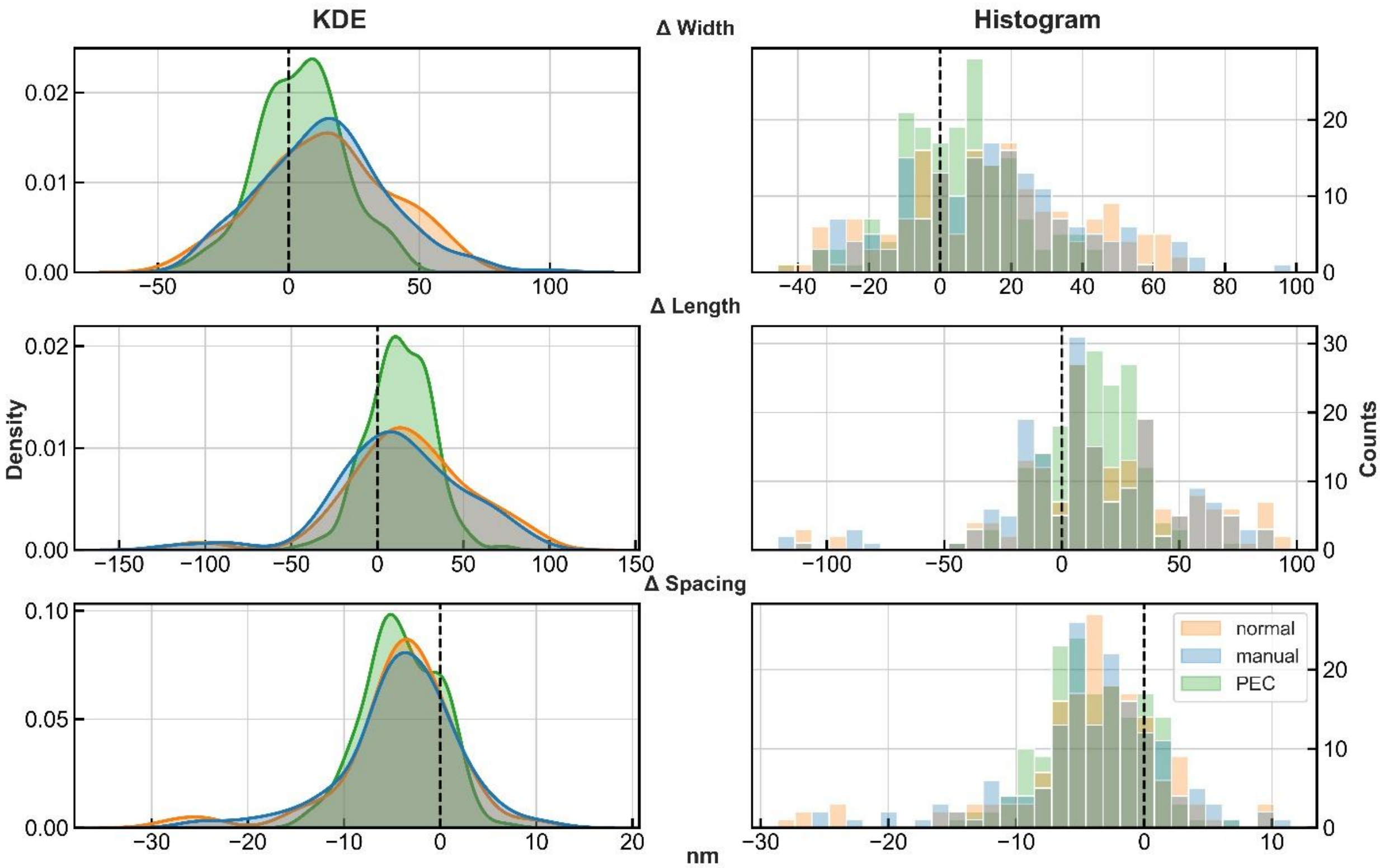

***Fig. S3*** *Statistical distribution of phononic shield deviations. Left: Kernel density estimation (KDE) plots of Δlength, Δwidth, and Δspacing for different fabrication conditions. Right: Corresponding histograms showing count distributions.*

Across all three metrics, the PEC-fabricated structures exhibit distributions most tightly centered around zero, indicating the greatest agreement with the intended design. The corresponding KDE curves are notably narrower and more sharply peaked than those of the other fabrication methods, reflecting reduced variability and improved uniformity. In contrast, normal fabrication exhibits broader distributions with more pronounced tails, indicating greater deviations and increased process-induced variability. The manual approach lies intermediate between the two, with moderate spread and slight offsets from zero.

These trends are consistently observed in both the KDE and histogram representations. The histogram counts further highlight the higher concentration of PEC data points near zero, whereas the normal process exhibits a wider dispersion across bins. Together, these results indicate that PEC provides the most accurate and precise pattern transfer, while the normal process introduces the greatest deviation from the design, with manual fabrication offering partial improvement.

## S9. Detection and Analysis of Photonic Cavity Holes

We analyzed the elliptical holes in the photonic cavity using OpenCV. We first identified hole candidates using a blob detection algorithm configured to detect dark, approximately circular regions within a specified area range. To isolate the cavity region, we retained only features located near the median vertical position. We removed closely spaced duplicate detections using a distance threshold. We sorted the remaining hole centers along the cavity axis.

For each detected hole, we extracted a local image patch and converted it into a binary image using Otsu thresholding, which determines an optimal threshold by minimizing intra-class variance between foreground and background. We selected the largest connected contour within the patch and fitted an ellipse using a least-squares approach. From this fitted ellipse, we obtained:

- major axis
- minor axis
- orientation

We computed effective horizontal and vertical radii ($r_x, r_y$) by accounting for ellipse rotation and converted them to nanometers using the calibration factor. We computed ellipticity as $r_y/r_x$ to quantify deviations from the intended shape. We also computed centre-to-centre distances between neighbouring holes along the cavity axis. To enable direct comparison with the reference design, we aligned the extracted hole positions with the GDS coordinates using the same procedure described earlier for the phononic shield structures (affine transformation, followed by nearest-neighbour matching using k-d tree). We overlaid the GDS hole geometries onto the SEM image to assess deviations in both position and shape relative to the intended design, providing visual validation of the alignment accuracy. (see Fig. 5a)

We quantified variations in hole geometry relative to the reference design by computing deviations in horizontal radius ($\Delta r_x$), vertical radius ($\Delta r_\gamma$), and inter-hole spacing (Δspacing). We extracted these quantities from precomputed datasets for each fabrication condition (normal, manual, and PEC). In Fig. 5b, we visualized these deviations as a function of the hole index along the cavity. For each fabrication condition, we generated line profiles with markers to enable direct comparison across corresponding holes.

We included a zero-reference line in each plot to indicate perfect agreement with the design. Positive and negative deviations correspond to over- and under-sizing relative to the GDS geometry, respectively. This representation provides a spatially resolved view of fabrication-induced variations along the cavity, allowing us to identify systematic trends, local deviations, and differences between exposure conditions.

**S10. Edge Deviation and Roughness Analysis**

We quantified edge deviations for both cavity holes and phononic shield features by comparing SEM-extracted contours with the corresponding reference geometries from the GDS layout. For each contour point, we mapped it relative to the ideal polygon and computed the signed radial deviation using a normalized distance function. We converted these deviations to nanometers and mean-centered them. We defined the total error as the root-mean-square (RMS) value of these deviations.

We expressed the deviation signal as a function of angular position about the feature center and applied a moving-average filter to extract the low-frequency component that represents systematic shape distortion. We defined the residual high-frequency component as line edge roughness (LER) and computed it as the standard deviation of the residual.

We computed the distortion component using:

$$\mathrm{RMS}^2 = \mathrm{LER}^2 + \mathrm{Distortion}^2$$

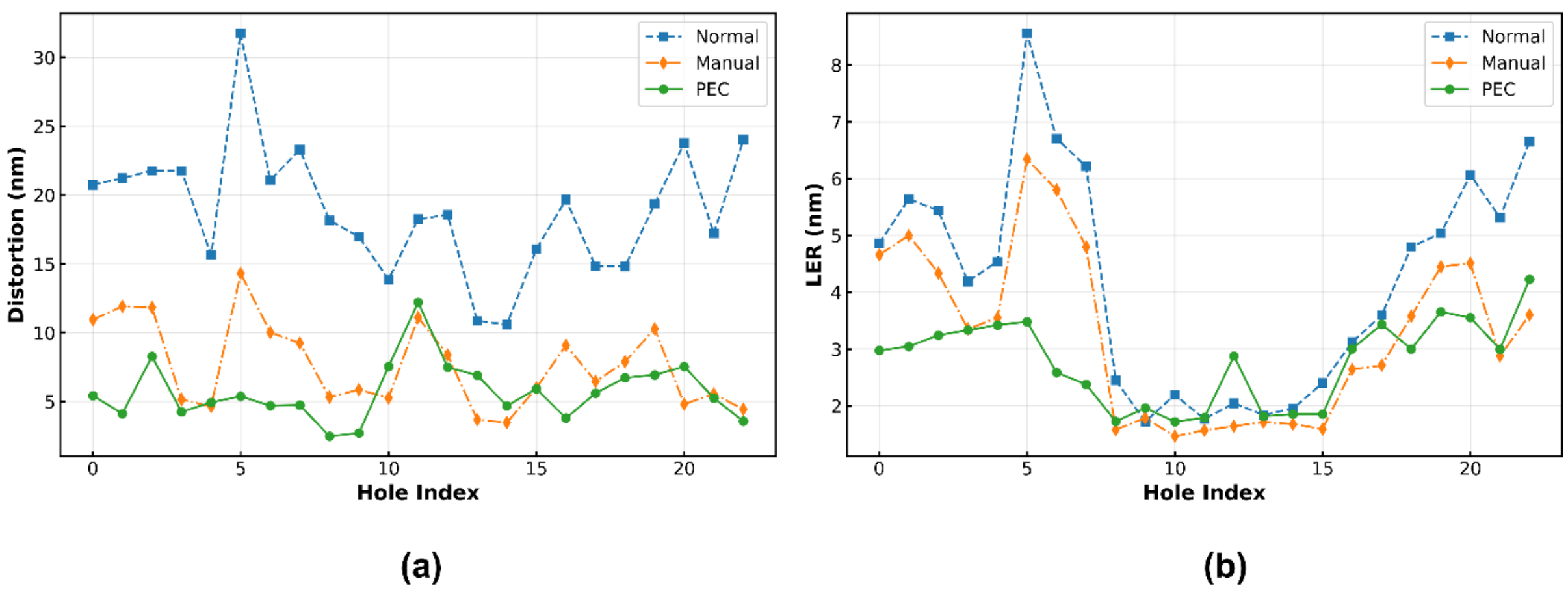

***Fig. S4*** *(a) Distortion and (b) LER vs hole index for different fabrication conditions.*

In **Fig. S4**, we show:

- **Fig. S4a**: Distortion vs hole index
- **Fig. S4b**: LER vs hole index for all fabrication conditions

## S11. Comparative Metrics Across Features

We computed mean values of all extracted parameters for both holes and plus structures.

In **Fig. S5**, we present:

- **Fig. S5a (holes)**: mean $\Delta r_x$, $\Delta r_y$, Δspacing, RMS, distortion, and LER
- **Fig. S5b (pluses)**: mean Δwidth, Δlength, Δspacing, RMS, distortion, and LER

for all SEM datasets.

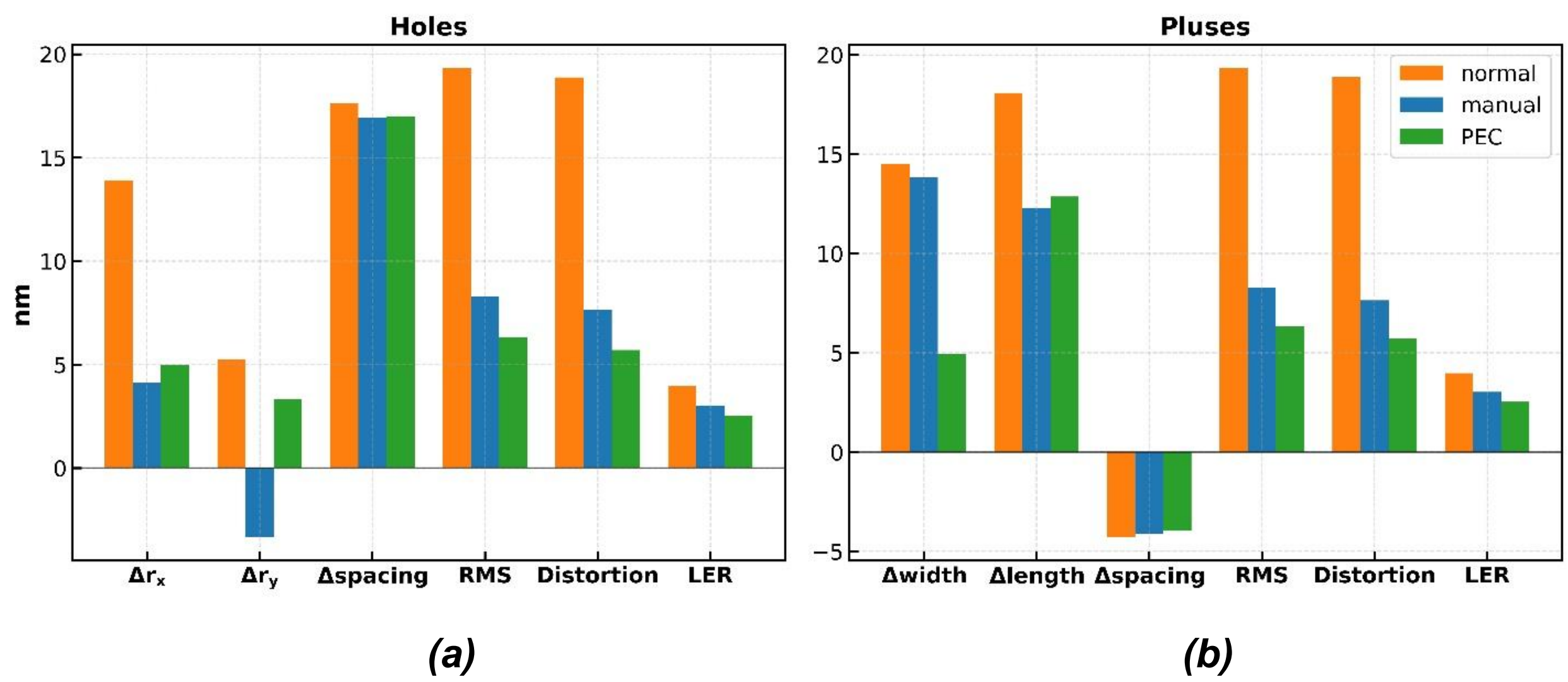

***Fig. S5*** *Mean geometric deviations and edge metrics for (a) cavity holes and (b) phononic shield features.*

These plots provide a compact comparison of fabrication performance across different exposure conditions.